\documentclass[10pt]{article}
\pdfoutput=1
\usepackage{jheppub}
\usepackage{amsmath,amssymb,amsfonts,mathbbol,graphicx,slashed,color,amsthm, mathtools, upgreek, enumerate, tensor}
\usepackage{subcaption}
\usepackage{setspace}
\usepackage[export]{adjustbox}
\usepackage{arydshln}
\usepackage[dvipsnames]{xcolor}
\usepackage{physics}
\usepackage[normalem]{ulem}
\usepackage{comment}

\graphicspath{{pics/}}

\allowdisplaybreaks

\colorlet{darkblue}{blue!70!black}
\colorlet{darkgreen}{green!50!black}
\colorlet{darkred}{red!50!black}

\newcommand{\mc}{\mathcal}
\renewcommand{\t}{\tilde}

\def\bea{\begin{eqnarray}}
\def\eea{\end{eqnarray}}
\def\be{\begin{equation}}
\def\ee{\end{equation}}

\title{
de Sitter Microstates from $T\bar T+\Lambda_2$ and the Hawking-Page Transition
}

 \author[a]{Evan Coleman,}
 \author[b]{Edward A. Mazenc,}
 \author[a]{Vasudev Shyam,}
 \author[a]{Eva Silverstein,}
 \author[a,c]{Ronak M Soni,}
 \author[d]{Gonzalo Torroba}
 \author[a]{and Sungyeon Yang}
 \affiliation[a]{Stanford Institute for Theoretical Physics, 382 Via Pueblo, Stanford CA 94305}
 \affiliation[b]{Enrico Fermi Institute \& Kadanoff Center for Theoretical Physics,
University of Chicago, Chicago, IL 60637, USA}
 \affiliation[c]{Department of Applied  Mathematics and Theoretical Physics,  University of Cambridge, Wilberforce Road, Cambridge, CB3 0WA, UK}
 \affiliation[d]{Centro At\'omico Bariloche and CONICET, S.C. de Bariloche, R\'io Negro, R8402AGP, Argentina}
 \vspace{5mm}

\vspace{1cm}

\emailAdd{ecol@stanford.edu}
\emailAdd{mazenc@uchicago.edu}
\emailAdd{evas@stanford.edu}
\emailAdd{vshyam@stanford.edu}
\emailAdd{rs2194@cam.ac.uk}
\emailAdd{torrobag@gmail.com}
\emailAdd{syang61@stanford.edu}

\abstract{

We obtain microstates accounting for the Gibbons-Hawking entropy in $dS_3$, along with a subleading logarithmic correction, from the solvable $T\bar T+\Lambda_2$ deformation of a seed CFT with sparse light spectrum.  The microstates arise as the dressed CFT states near dimension $\Delta=c/6$, associated with the Hawking-Page transition; they dominate the real spectrum of the deformed theory.  We exhibit an analogue of the Hawking-Page transition in de Sitter.  Appropriate generalizations of the $T\bar T+\Lambda_2$ deformation are required to treat model-dependent local bulk physics (subleading at large central charge) and higher dimensions. These results add considerably to the already strong motivation for the continued pursuit of such generalizations along with a more complete characterization of $T\bar T$ type theories, building from existing results in these directions.  

}

\begin{document}

\maketitle
\parskip=10pt

\section{Introduction:  from Bekenstein-Hawking to Gibbons-Hawking via Hawking-Page} \label{sec:intro}

It is of considerable interest to formulate quantum gravity consistently with the observed accelerated expansion of the universe.  Recent years have seen renewed progress toward this goal stemming in part from a simple  generalization \cite{GST, LLST} of the $T\bar T$ deformation \cite{Zamolodchikov:2004ce, Smirnov:2016lqw, Dubovsky:2012wk, Cavaglia:2016oda, Dubovsky:2017cnj, Gorbenko:2018oov} applied to AdS/CFT duality, as pioneered in \cite{McGough:2016lol, Kraus:2018xrn, Hartman:2018tkw}.  
In this correspondence, patches of AdS or dS spacetime are reconstructed from a deformed holographic seed CFT  formulated on the boundary of a patch; energy levels and various entanglement entropies accessible in the gravitational sector agree precisely on the two sides \cite{Donnelly:2018, Donnelly:2018bef, GST, LLST}.  The global cosmological spacetime may be obtained by sewing together such building blocks.   This led to a recent calculation of the Gibbons-Hawking entropy $S_{GH}$ in \cite{Shyam:2021ciy}, reproducing \cite{Dong:2018cuv}\ and a logarithmic correction computed recently in \cite{Anninos:2020hfj}.  
\begin{equation}\label{eq:Scorrected}
    S=S_{GH}-3 \log (S_{GH})+\dots
\end{equation}

In the present work, we define a version of this deformation which provides a microstate count for the de Sitter observer patch, reproducing the corrected Gibbons-Hawking entropy \cite{Gibbons:1977mu}\ \eqref{eq:Scorrected} in a more direct way. Specifically, we construct microstates of the de Sitter observer patch from the microstates of a particular black hole in AdS.
The generalization \cite{GST, LLST} of the $T\bar T$ deformation accommodating de Sitter spacetime (known as the $T\bar T+\Lambda_2$ deformation) is defined by 
\begin{equation}\label{eq:ZTTbLambda}
\frac{\partial}{\partial \lambda}\, \log Z =-2\pi \,\int d^2 x\,\sqrt{g}\, \langle T \bar T \rangle+\frac{1-\eta}{2\pi \lambda^2}\int d^2 x \,\sqrt{g}
\end{equation}
with $Z$ the partition function of the theory and where $T\bar T$ is defined as the quadratic combination
\begin{equation}\label{eq:TTbdef}
    T\bar T \equiv \frac{1}{8}(T_{ab}T^{ab}-(T^a_a)^2)
\end{equation}
of components of the energy-momentum tensor $T_{ab}$.  

\begin{figure}[t!]
  \centering
  \includegraphics[width=0.95\linewidth]{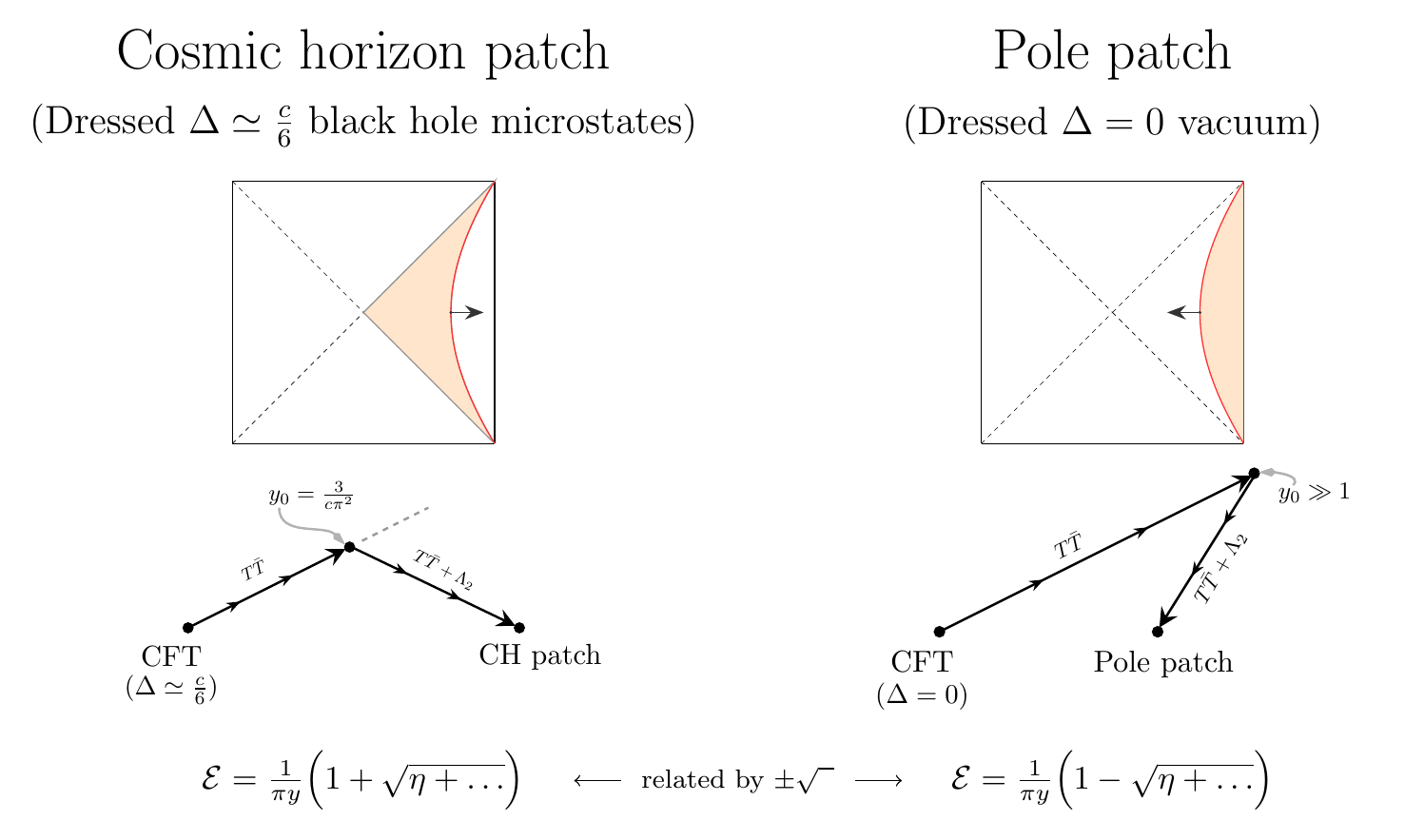}
  \caption{ \small \sffamily 
  A schematic of our prescription for holographic reconstruction of the static patch, including a microstate count, for a given deformation parameter $y=\lambda/L^2$.  This may be summarized as follows, with details in the bulk of the paper.  On the left, we obtain the indicated patch containing the cosmic horizon by dressing the $\Delta\simeq c/6$ microstates comprising a BTZ black hole at the Hawking-Page transition and switching between the $T\bar T$ and $T\bar T+\Lambda_2$ trajectories at a value $y_0$ of the deformation parameter such that the BTZ and Cosmic Horizons degenerate to indistinguishable near-horizon regions.  On the right we depict the trajectory described in \cite{GST}\cite{LLST} which formulates the complementary patch for the same boundary geometry.  The extrinsic curvature of the boundary, and hence the square root term in the dressed energy formula, are equal and opposite for the two patches.  As contributions to the thermal partition function at fixed boundary geometry, these dominate for complementary ratios of $(\beta/L)^2=y/y_0-1$, providing a de Sitter analogue of the Hawking-Page transition.  In both cases, we can capture the bulk of the static patch by continuing the trajectory in the manner indicated by the arrow.  The Pole patch with a Dirichlet boundary condition always excludes the cosmic horizon and does not directly account for the microstates, in keeping with the sparse set of real dressed energies in that case. However, for the cosmic horizon patch -- the main subject of the current paper -- proceeding with the trajectory to $y_{final} \gg 1$ formulates the static patch with boundary at the observer position at the North Pole. }
  \label{fig:3dfriendly}
\end{figure}

The prescription \eqref{eq:ZTTbLambda} is applied piecewise, beginning with a pure $T\bar T$ deformation ($\eta=1$) from the seed CFT theory.  Once $\lambda$ is nonzero, it is possible to start a new segment of the trajectory with nonzero $\eta-1$ beginning at some finite value $\lambda=\lambda_0$ (say taking $\eta=-1$), as sketched in figure \ref{fig:3dfriendly}.
This is universal and solvable in the same sense as the original $T\bar T$ theory, enabling a calculation of the deformed (`dressed') energy spectrum for all energy levels.

In our new example of such a combined $T\bar T\to T\bar T+\Lambda_2$ deformation, the microstates associated with the de Sitter observer patch are dressed microstates of BTZ black holes at the $\Delta \simeq c/6$ energy level of the seed CFT of central charge $c\gg 1$ ($\Delta$ denoting the operator dimension).\footnote{As we will describe in the bulk of the paper, the precise statement of the relevant energy levels is as a band of order 1 around $\Delta =c/6$ \cite{Hartman_2014, Mukhametzhanov:2019pzy}. } In the dual gravitational description, the trajectory corresponds to bringing in the AdS boundary until it skirts the horizon of the $\Delta \simeq c/6$ BTZ black hole, where the horizon becomes indistinguishable from the horizon of the dS static patch.  At this value of $\lambda=\lambda_0$, we adjoin a $T\bar T+\Lambda_2$ segment of the trajectory, building up the static patch of dS, bounded by a worldline at its observer position; see Fig.~\ref{fig:3dfriendly}.\footnote{For earlier work on holography and the observer patch see e.g. \cite{Anninos:2011af}.} Throughout this process, the entire spectrum of microstates accounting for the Gibbons-Hawking entropy are retained as real energy levels.  
This microcanonical state count has a counterpart 
in the canonical ensemble, where the $\Delta=c/6$ energy level corresponds to the Hawking-Page transition in the seed AdS/CFT system. In the course of our analysis below, we find an analogue of the Hawking-Page transition in de Sitter.  

This derivation of de Sitter microstates is our main result, recovering the connection between the Gibbons-Hawking entropy and the $\Delta=c/6$ states associated with the Hawking-Page transition that was anticipated in \cite{Dong:2018cuv}\ and recovered with the first $1/c$ correction in \cite{Shyam:2021ciy}.  As we will see, there are different versions of such a trajectory with distinct properties, having to do with choices of integration constants arising in the differential equation for the dressed energy levels derived from \eqref{eq:ZTTbLambda}, related to the role (or not) of modular transformations in the theory.  In holography, the prescription \eqref{eq:ZTTbLambda}\ captures the pure gravitational sector of the theory -- the black hole spectrum in AdS and the observer patch in dS. The physics of other energy levels is subdominant at large $c$ and model-dependent, as we will describe in our analysis below.   A complete description of subleading excitations (such as local bulk matter) requires additions to the prescription \eqref{eq:ZTTbLambda} including those analyzed for bulk scalar fields in \cite{Hartman:2018tkw}.    
We will also explain the relation of the present results with those of \cite{GST, LLST, Shyam:2021ciy}.

\section{$T\bar T+\Lambda_2$ trajectories, patchwise holography, and microstates} \label{sec:review}

In the pure gravity sector, finite-cutoff holography in $AdS_{3}$ and $dS_{3}$ has recently been shown, starting with the work of \cite{McGough:2016lol,Kraus:2018xrn,GST} to be dual to certain deformations of a holographic CFT.
Although cosmological spacetimes do not contain time-like boundaries,\footnote{With the exception of specific UV-complete solutions with end-of-the-world codimension-one orientifold planes such as \cite{Polchinski:1998rq} and relatives such as \cite{Horava:1996ma}.} cut-off patches of spacetime are useful as building blocks for a formulation of the global spacetime (the full spacetime being obtained by sewing such patches together with an integral over their bounding geometry).\footnote{For a different approach involving finite patches of spacetime see e.g. \cite{Banks:2018ypk}.}  Moreover, in the special case that the finite patch is the domain accessible to an observer, bounded by their trajectory and conditioned on their existence and experience,\footnote{We thank Zhenbin Yang for suggesting this perspective in related discussions. The fixed metric condition at the Dirichlet wall below corresponds to conditioning on the metric experienced by this observer.}  the dual formulation may provide more than just a building block. The prescriptions \cite{McGough:2016lol, Kraus:2018xrn, GST, LLST} along with the new versions we introduce in the present work provide concrete holographic duals for finite patches,  applicable to both situations.
In this section, we give a quick review of the connection, generalizing it appropriately for our present purposes. 

At the end of this section, we will get to the advertised microstate construction obtained by judicious choice of $T\bar T+\Lambda_2$ trajectory.  We will focus on the dressing of the $\Delta \simeq c/6$ band of energy levels to the de Sitter ground state.  The role of other levels depends on additional specifications in the trajectory, which we will elucidate for two versions in the subsequent sections \S\ref{sec:3d-friendly}-\ref{sec:2d-friendly}.  In fact, as we will see those prescriptions essentially reduce to the micro-states we identify in the present section.

\subsection{General formulation}
A $T \bar{T}$-deformed CFT is defined by the differential equation, and boundary condition,
\begin{equation}
  \frac{\partial}{\partial \lambda}\, \log Z =-2\pi \,\int d^2 x\,\sqrt{g}\, \langle T \bar T \rangle+\frac{1-\eta}{2\pi \lambda^2}\int d^2 x \,\sqrt{g}\;, \qquad Z_{\lambda = 0, \eta=1} = Z_{CFT},
  \label{eqn:ttbar-eqn}
\end{equation}
where $Z(\lambda, g)$ is the partition function of the theory on a manifold with metric $g$, the $T \bar{T}$ operator is the composite of the stress tensor defined above in \eqref{eq:TTbdef}, and \eqref{eqn:ttbar-eqn} is expressed in Euclidean signature.\footnote{Appendix \ref{sec:notationconventions} collects our notation and conventions.}   We will be particularly interested in the solvable case where the theory is formulated on a flat space, with $Z$ depending on the circumference $L$ of the cylinder; in the case of the canonical ensemble at finite temperature it depends on both $L$ and the Euclidean time period $\beta$ (in addition to its dependence on the deformation parameter $\lambda$).
We sometimes denote the second term here as $\Lambda_2$ (distinct from the bulk $3d$ cosmological constant $\Lambda_3$ whose sign will correspond to $-$sign($\eta$)). $Z_{CFT}$ is the partition function of the `seed' CFT that we deform by this procedure.  As we will explain shortly, one can define a deformed theory by joining together solutions to this equation with different values of $\eta$, given appropriate boundary conditions on each segment of the trajectory. 

This deformation is most rigorously defined for all QFTs on flat two-dimensional manifolds \cite{Zamolodchikov:2004ce,Cardy:2018sdv}, though progress has been made on curved spaces \cite{GST, Mazenc:2019cfg} and higher dimensions \cite{Hartman:2018tkw} in perturbation theory.
Here, we work only with holographic seed CFTs, where there is a large-$c$ and large 't Hooft coupling semi-classical limit, for which the theory has additional bulk locality properties and enhanced theoretical control.  This entails a sparse light spectrum in the sense of \cite{Hartman_2014, Mukhametzhanov:2019pzy}, which guarantees a Cardy entropy count beginning with the energy level corresponding to operator dimension $\Delta = c/6$.   We will be most interested in universal aspects of the physics, captured by the Einstein gravitational sector which well describes black hole solutions in AdS and the de Sitter cosmic horizon.  At the same time, we will keep track of where this description becomes inadequate, as it does for energy levels arising from model-dependent particle states in the sparse low-energy part of the seed CFT spectrum.

As indicated in \eqref{eqn:ttbar-eqn}, a differential equation is not sufficient to define a theory; we also need a boundary condition for each segment of the trajectory.
For $\eta = 1$, when the second term vanishes, there is an obvious boundary condition at $\lambda = 0$,
\begin{equation}
  Z_{\eta=1,\lambda} \xrightarrow{\lambda \to 0} Z_{CFT}.
  \label{eqn:ads-def-bd-cond}
\end{equation}
Once $\lambda$ is nonzero, we may split off a trajectory including the $\Lambda_2$ deformation with $\eta=-1$ starting from any nonzero value $\lambda_0$.  

Before continuing, it is useful to incorporate the fact that the physics depends only on the appropriate  dimensionless quantities.  The deformed path integral has the scale-invariance 
\begin{equation}
  Z [\lambda,\lambda_0, g_{2}] = Z \left[ e^{2 \Omega} \lambda, e^{2\Omega}\lambda_0, e^{2 \Omega} g_{2} \right],
  \label{eqn:scale-invariance}
\end{equation}
where $\Omega$ is a constant.
We will mainly be interested in the case when the manifold is a cylinder with circumference $L$ (or a torus with cycle lengths $\beta, L$ in some Euclidean calculations).
Let us introduce the shorthand notation
\begin{equation}\label{eq:ydef}
    y=\frac{\lambda}{L^2}, \quad y_0 = \frac{\lambda_0}{L^2}\,.
\end{equation}
It is useful to think of the trajectory as obtained by varying the dimensionless variable $y$.  As we will review shortly, the gravity side holographically starts by bringing in the boundary, most directly changing $L$ \cite{McGough:2016lol, Kraus:2018xrn, GST, LLST}, while the description natural in the 2d language involves changing $\lambda$ as in the defining equation \eqref{eqn:ttbar-eqn} while working on a fixed cylinder; in both descriptions $y$ changes along the trajectory.    

Given a nonzero value of $y_0$, we have a boundary condition
\begin{equation}\label{eqn:general-bc}
   Z_{\eta \neq 1, y = y_0} =  Z_{\eta = 1, y=y_0}~~~~{\text{(general~case)}}\,.
\end{equation}
In saying this, we must keep in mind that the simple formulation as written in \eqref{eqn:ttbar-eqn}\ pertains holographically only to universal (`pure gravity') quantities in the theory.  Below we will describe in detail the implementation of the boundary conditions at the level of the appropriate energy levels in the system.

Previous work \cite{GST, LLST} focused on the case where we begin this $T\bar T+\Lambda_2$ trajectory at $y_0 \gg 1$.
The $\Lambda_{2}$ term disappears at $\lambda \to \infty$, and we may set a boundary condition in this limit
\begin{equation}
  \lim_{\lambda \to \infty} Z_{\eta \neq 1, \lambda} = \lim_{\lambda \to \infty} Z_{\eta = 1, \lambda} ~~~~{\text{(previous~work)}}\,.
  \label{eqn:other-def-bd-cond}
\end{equation}
In the present work, we will introduce a different deformed theory in which the $T\bar T+\Lambda_2$ trajectory splits off from the pure $T\bar T$ trajectory well before $y=\lambda/L^2$ reaches infinity. As motivated above in the introduction, this will yield a microstate count for the de Sitter static patch by dressing CFT states at dimension $\Delta =h+\bar h=c/6$ to the $dS_3$ static patch, obtained via a splitting off point $y_0=3/\pi c^2$. 

As pioneered in \cite{Zamolodchikov:2004ce, Smirnov:2016lqw, Cavaglia:2016oda}, the dressed energy levels comprise a key computable quantity in the deformed theory.\footnote{See e.g. \cite{Cardy:2019qao} and \cite{Kruthoff:2020hsi} for interesting studies of the behavior of additional quantities (correlation functions and wavefunctions) under the deformation.}  The original derivation proceeds by expressing the deformation in terms of the interaction Hamiltonian, deriving a differential equation for the energy levels (applied level by level). Equivalent results arise from an inverse Laplace transform of the partition function as in \cite{Shyam:2021ciy} and from the $2d$ gravity reformulation of the theory \cite{Dubovsky:2017cnj,Dubovsky:2018bmo, Mazenc:2019cfg}.  With $\Lambda_2\propto 1-\eta$ included, the differential equation for the energy levels becomes (see Appendix A of \cite{GST})
\be\label{eq:Energy-diffeq}
\pi y \mc E(y) \mc E'(y)- \mc E'(y)+ \frac{\pi}{2} \mc E(y)^2=  \frac{1-\eta }{2 \pi  y^2}+2\pi^3 J^2
\ee
where we defined the dimensionless dressed energy
\begin{equation}\label{eq:calE}
    {\cal E} = E L\,.
\end{equation}
For a seed CFT,
\be
{\cal E}\big|_{y=0, \eta=1}={\cal E}_{CFT}=2\pi\left(\Delta -\frac{c}{12}\right)\,.
\ee
Again here, a given $T\bar T+\Lambda_2$ theory will entail particular boundary conditions on this equation.

The general solution to the differential equation takes the form
\be\label{eqn:S1-En-gen}
\mc E(y) = \frac{1}{\pi y} \left(1 \pm \sqrt{\eta-4 C_1 y+4 \pi^4 J^2 y^2} \right)\,,
\ee
with the integration constant $C_1$ and the branch of the square root determined appropriately for each segment of the trajectory (meaning the initial segment consisting of the $T\bar T$ deformation alone, and the final segment deforming by $T\bar T+\Lambda_2$ starting from $y_0$).
Here $J$ is the spin; in terms of left and right scaling dimension $(h, \t h)$, we have $\Delta= h + \t h$, $J= h-\t h$.

For quantities admitting a smooth `pure gravity' description -- which are the only ones captured by the pure $T\bar T$ and $T\bar T+\Lambda_2$ deformations -- we will require continuity in the energy at $y=y_0$ (in addition to the boundary condition matching the seed at $y=0$ \eqref{eq:Energy-diffeq}).  We do not generally require continuity of this formula for all energy levels, for the following reason.  In general, in order to capture local bulk matter in detail,  the prescription for the deformation \eqref{eq:ZTTbLambda} will require additional model-dependent specifications such as those carried out in the scalar sector in \cite{Hartman:2018tkw}.  For such model-dependent contributions to the spectrum, subdominant for $c\gg 1$,  the differential equation for the energies \eqref{eq:Energy-diffeq} and its solutions \eqref{eqn:S1-En-gen}\ require additional contributions; it is only those corrected formulas that must be continuous.
Happily,  the entropically dominant states -- those of our main interest -- are captured by the solvable and universal pure $T\bar T+\Lambda_2$ deformation.

In the remainder of this section, we will  focus on this universal piece of the physics, associated with the observer patch of the $dS_3$ vacuum as dressed $\Delta =c/6$ BTZ black holes.   In subsequent sections, we will lay out separate versions of the complete theory, accounting for the fate of all energy levels (with due regard to the appropriate level of model-dependence in $dS_3$ holography).

\subsection{{Gravity side and dictionary}}

At this point, it is useful to review the basic idea of the map to the gravity side in the holographic correspondence developed in \cite{McGough:2016lol, Kraus:2018xrn, GST, LLST}.  
The differential equation for the energy levels is equivalent to the Einstein equations in a bounded patch of (A)dS spacetime.  This connection proceeds starting from the bulk gravity action (in Lorentzian signature) 
\begin{align} \label{eqn:lorentz-grav}
I &=  \frac{1}{16\pi G_N} \int_{\mathcal{M}_3} d^3 x \sqrt{-g} \left( R + \frac{2}{\ell^2}\eta \right) + \frac{1}{8\pi G_N} \int_{\partial\mathcal{M}_3} d^2 x \sqrt{-g} \left(K-\frac{1}{\ell} \right) 
\end{align}
with boundary condition 
\begin{equation}
 ds^2_{\mathcal{M}_3}\Big|_{\partial\mathcal{M}_3} =  
 ds^2_{\mathcal{M}_2}\,.
\end{equation}

Specializing to the case where the boundary metric is a flat cylinder of proper circumference $L$, the differential equation for the energy given above in \eqref{eq:Energy-diffeq} follows from the Einstein equations \cite{McGough:2016lol, Kraus:2018xrn}, once we identify the energy with the Brown-York energy \cite{Brown:1977sj}.\footnote{with similar statements holding for the curved case perturbatively around large $c$} The Einstein equations can be processed as reviewed in the appendix of \cite{GST} to yield the equation \eqref{eq:Energy-diffeq} for 
\begin{equation}
    {\cal E} = L\int_0^L dx \, T^t_t
\end{equation}
in terms of cylinder coordinates $-dt^2+dx^2$, with the
Brown-York stress-energy tensor given by
\begin{equation}
  T_{\mu\nu} = { \frac{2}{\sqrt{-g}}\frac{\delta I_\text{on-shell}}{\delta g^{\mu\nu}}}=\frac{1}{8 \pi G_N} \left( K_{\mu\nu} - g_{\mu\nu} K + \frac{1}{\ell} g_{\mu\nu} \right)\,
  \label{eqn:BY-formula}
\end{equation}
in terms of the extrinsic curvature $K_{\mu\nu}$ of the Dirichlet wall. In this correspondence, the map of parameters is
\begin{equation}\label{eq:param-dictionary}
    c=\frac{3\ell}{2 G_N}, ~~~ \lambda=8 G_N\ell\,.
\end{equation}

In this correspondence, different choices of branch and integration constant in the solutions \eqref{eqn:S1-En-gen} correspond to different choices of the patch and its excitations (insofar as they are contained in the `pure gravity' sector of the theory).  This will be a key technical point of our work, which we will spell out in detail in what follows for the cases of interest.  In particular, the choice of branch of the square root in \eqref{eqn:S1-En-gen} corresponds to the sign of the extrinsic curvature $K$.  This is not immediately obvious from inspection, but can be seen from the derivations in \cite{McGough:2016lol, Kraus:2018xrn, GST} (in which the non-square-root term is tied to a relatively inconsequential counterterm).\footnote{This counterterm is chosen first of all to match the one required for AdS/CFT \cite{Balasubramanian:1999re} in the limit $y\to 0$, and for simplicity we maintain the same value for it throughout the trajectories we consider here.}  As such, opposite signs of the square root correspond to complementary ways to fill in the bulk patch starting from the same Dirichlet wall.  To anticipate a little, this is the difference between the patches depicted in figure \ref{fig:3dfriendly}.      

In order to capture a particular bulk patch as a gravitational representation of a dressed energy level in this way, we must specify the full trajectory including the choices of branch and integration constant. At this point, rather than continuing with the generalities, we will shortly spell this out for several cases of interest.

To proceed to that, let us record metrics for AdS and dS vacua and for BTZ black holes (with these and particle states related by various identifications).    
First, the locally-$AdS_{3}$ metric
\begin{equation}
  ds_{3}^{2} = - \left(\frac{r^{2} - r_{h}^{2}}{l^{2}}\right) d\tau^{2} + \left(\frac{l^{2} }{r^{2} - r_{h}^{2}}\right)dr^{2} + r^{2} d\phi^{2}, \qquad r_{h}^{2} \ge -\ell^2,\ \phi \sim \phi + 2\pi
  \label{eqn:3d-g-ads}
\end{equation}
captures the $AdS_3$ global vacuum and spinless BTZ black holes. 
The case $r_{h}^{2} = -\ell^2$ is global $AdS$.
In terms of the black hole mass, $8G_N M = r_h^2/\ell^2$, we have $\ell M = \Delta-c/12$, leading to the dictionary (adding to \eqref{eq:param-dictionary} above)
\begin{equation}\label{eq:Delta-dictionary}
    \Delta- \frac{c}{12}= \frac{r_h^2}{8 G_N \ell} = \frac{1}{2\pi}{\cal E}_{y=0} 
\end{equation}
where the last subscript indicates the dimensionless energy ${\cal E}=EL$ in the seed CFT.    
As mentioned above, the spectrum of the holographic seed theory breaks up into the following windows (see for instance \cite{Hartman_2014, Mukhametzhanov:2019pzy}):
\bea\label{eq:CFT-windows}
\Delta &<& \frac{c}{12}:~  {\text{sparse~spectrum ~(particle~states)}} \nonumber\\
\Delta &\ge& \frac{c}{6}:  S\simeq S_{Cardy}= 2\pi\sqrt{\frac{c}{3}\left(\Delta -\frac{c}{12}\right)}  
\eea
(in between which is a window of light BTZ black hole states). More precisely, the detailed CFT states are generically non-degenerate, split by energy differences of order $\exp(-S)$.  This enables a continuum approximation to the density of states; in recent work, the count of states given by $S_{Cardy}$ \cite{Mukhametzhanov:2019pzy} was shown to apply to an order one band (i.e. order $c^0$) around a fixed $\Delta \sim {\cal O}(c)$ level. 
We note from \eqref{eq:param-dictionary}, \eqref{eq:Delta-dictionary} a relation which will play a role in our analysis (following \cite{Dong:2018cuv}):
\begin{equation}\label{eq:rh-ell-HP}
    \Delta=\frac{c}{6}\,\, \implies \,\, r_h=\ell\,.
\end{equation}
We will see below in \S\ref{sec:HP-mod} that this energy level has a special place in the spectrum associated with modular transformations and the Hawking-Page transition. These results are general; one may consider as an example of a seed theory the D1-D5 CFT, whose black hole entropy count is known microphysically \cite{Strominger:1996sh}\ and for which an uplift to de Sitter was constructed in \cite{Dong:2010pm}.

Secondly, the static patch metric for $dS_3$ is
\begin{equation}
  ds_{3}^{2} = - \left(\frac{\ell^2 - r^{2} }{{\ell^2}}\right) d\tau^{2} + \left(\frac{{\ell^2} }{\ell^2-r^{2}}\right)dr^{2} + r^{2} d\phi^2\,.
  \label{eqn:static-patch}
\end{equation}
In both cases, various further identifications yield additional states; e.g. restricting the period of $\phi$ in \eqref{eqn:static-patch}\ describes the metric sourced by a particle at the pole at $r=0$ \cite{GST}, and more elaborate identifications introduce spinning states (see e.g. \cite{Bousso:2001mw}).

\subsection{{ Trajectories and horizon entropy}}

Now that we have laid out the formalism on both sides of the duality, let us explain the key examples of interest here.  In brief:  we will review how the original version of the $T\bar T+\Lambda_2$ trajectory captures a patch of spacetime not containing the cosmic horizon but containing the observer trajectory (say the North Pole) within the static patch.  The new version captures the complementary part of the static patch, including the cosmic horizon.  The latter exhibits the microstates (with the correct count) as appropriate dressed energy levels of the seed CFT.

\subsubsection{Previous Version:  The Pole Patch}\label{subsec:previous}

First we will quickly review the original prescription carried out in \cite{GST, LLST}.  There, we work with the boundary condition \eqref{eqn:other-def-bd-cond}\ for the $T\bar T+\Lambda_2$ part of the trajectory, meaning we continue with the pure $T\bar T$ trajectory until $y=\lambda/L^2$ reaches the matching point $y_0\gg 1$.  This first segment of the trajectory, applied to the ground state, entails bringing in the boundary of global $AdS_3$ -- prescribing a Dirichlet wall at $r=r_c=L/2\pi$ and filling in the region $r<r_c$  until the bounding cylinder is very small, with the patch inside of it being indistinguishable from a similar patch centered on an observer worldline (e.g.the North Pole $r=0$) in de Sitter spacetime.  Then beginning the $\eta=-1$ part of the trajectory, with the appropriate choice of branch and integration constants,  formulates a patch of $dS_3$ with $r\le r_c$.  
The energy formula for this case is given by 
\begin{equation}\label{eq:Eold}
    {\cal E}=\frac{L^2}{\pi\lambda} \left(1- \sqrt{\eta - \frac{4\pi^2\lambda}{L^2}\left(\Delta - \frac{c}{12}\right) + 4\pi^4 J^2 \frac{\lambda^2}{L^4}}\right)\,.
\end{equation}
This produces the correct CFT energy at $y=\lambda/L^2\to 0$, $\eta=1$, and at the matching between $T\bar T$ and $T\bar T+\Lambda_2$ the energies are continuous since $y=\lambda/L^2\to\infty$.

Note that the gravity-side patch that is thus formulated does not contain the cosmic horizon at $r=\ell$.  For $r_c\to \ell$, $\Delta=0$ this (almost) captures the static patch of de Sitter, but without the cosmic horizon given this order of limits, a distinction that will greatly concern us in this work.     
What this corresponds to in the deformed-CFT dual is the following: 
in the family of (A)dS patches we just reviewed that were formulated in the original trajectory \cite{GST}\cite{LLST}, only the particle states \eqref{eq:CFT-windows}\ survive in the real spectrum.  This is not enough to account directly for the Gibbons-Hawking microstates.  That fact agrees with the absence of the cosmic horizon (CH) in the patch.        

We are now ready to present the essence of the new trajectory, one that formulates the complementary patch containing the cosmic horizon, with a concommitant count of microstates.

\subsubsection{New Version: The Cosmic Horizon (CH) Patch} \label{subsec:new}
The branching-off point $y_0$ for the $T\bar T + \Lambda_2$ part of the trajectory may be chosen differently from that in the previous section, so long as the energy levels that are accessible in the model-independent pure-gravity sector of the theory are continuous in the transition from $\eta=1$ ($\Lambda_3<0$) to $\eta=-1$ ($\Lambda_3>0$).  
Moreover, as we stressed above in \eqref{eqn:S1-En-gen}, the solution space of the differential equation for the energies has choices of integration constant $C_1$ and a choice of the branch of the square root (in principle, these could be chosen separately level by level).

In this section, we will see that this freedom enables the construction of a consistent class of trajectories which formulate the cosmic horizon (CH) region of the static patch as a set of dressed states in the band of energies around $\Delta=c/6$.  At the end of the trajectory, this captures the entire static patch, bounded by the observer worldline at (say) the North Pole of the bulk de Sitter spacetime.

Let us begin with a standard observation --- the similarity between the near-horizon regions in a black hole spacetime \eqref{eqn:3d-g-ads} with $r_{h} = l$ and the $dS_{3}$ static patch \eqref{eqn:static-patch} with the same curvature radius $\ell$.
In the black hole case, we make the coordinate transformation 
\begin{align}
  r \equiv r_h + \delta r = \ell+\delta r, \qquad \delta r = \frac{w^2}{2\ell} \ll r_h = \ell,
  \label{eqn:bh-coord}
\end{align}
to find that the black hole metric \eqref{eqn:3d-g-ads} gets transformed into a nearly flat near-horizon region 
\begin{equation}
  ds_{3}^{2} = - \frac{w^2}{\ell^2} d\tau^{2} + d w^2 + \ell^2 d\phi^{2} +\dots
  \label{eqn:bh-nh}
\end{equation}
Similarly, in the $dS_{3}$ vacuum we take
\begin{equation}
  r \equiv \ell - \delta r, \qquad \delta r = \frac{w^2}{2\ell} \ll \ell \,.
  \label{eqn:ch-nh}
\end{equation}
Keeping one more contribution, for both cases the near-horizon metric takes the form
\begin{equation}\label{eq:nh-next-wsqrd-term}
ds_{3}^{2} = - \frac{w^2}{\ell^2} d\tau^{2} + d w^2 + (\ell^2 + 
\eta w^2) d\phi^{2}.
\end{equation}
where as usual $\eta=1$ for the $r_h=\ell$ BTZ case and $\eta=-1$ for the $dS_3$ vacuum.
Thus we find find exactly the same near-horizon metric \eqref{eqn:bh-nh}\ at leading order in an expansion in $\delta r=w^2/2\ell$.
Note that the value $r_h=\ell$ at which these match corresponds to the energy level $\Delta=c/6$ \eqref{eq:rh-ell-HP} \cite{Dong:2018cuv}.    

Recall that the bulk intuition for the boundary condition \eqref{eqn:other-def-bd-cond} was precisely that the spacetime near $r=0$ is nearly flat in either global $AdS_3$ \eqref{eqn:3d-g-ads} or the static patch of $dS_3$ \eqref{eqn:static-patch}, i.e. it becomes independent of the bulk cosmological constant in the relevant limit; in that case the system reverted to a nearly flat bulk spacetime contained within a small bounding cylinder.  It  is quite analogous that the near-horizon region boils down similarly to a nearly flat patch of spacetime.  The extrinsic curvature $K_{\mu\nu}$ and hence the Brown-York stress energy \eqref{eqn:BY-formula} agree at leading order in this approximation (when we drop the $\eta$-dependent term in \eqref{eq:nh-next-wsqrd-term}).

In particular, the extrinsic curvature of the metric \eqref{eq:nh-next-wsqrd-term}\ approaches zero in this region:
\be\label{eq:nh-K-tozero}
K_{\mu\nu}=\frac{1}{2}\partial_w g_{\mu\nu} \sim {\cal O}(w)\to 0
\ee
This non-trivial agreement at leading order in $w$ suggests that there is a new class of trajectories that may be defined, with a consistent boundary condition at a finite value of $y_0$, in which the transition from $\eta=1$ to $\eta=-1$ occurs when the system formulates equivalently the near-horizon region of either the $\Delta =c/6$ black hole or the $dS_3$ static patch.

As explained above, the square root term in the energy formula (given in general form in \eqref{eqn:S1-En-gen}) is proportional to the extrinsic curvature of the boundary on the gravity side. The matching of the two segments of the trajectory, at the indistinguishable horizon on the gravity side, should occur for the value of $y_0$ where this square root term vanishes.  At that point, we are free to continue with the $T\bar T+\Lambda_2$ trajectory with either sign of the square root, still maintaining continuity of the energy.   

Indeed, if we choose 
\begin{equation}\label{eq:y0}
    y_0 = \frac{3}{c\pi^2}
\end{equation}
then we see that the square root term in the $\eta=1$ (pure $T\bar T$) energy formula \eqref{eq:Eold} vanishes at $y\simeq y_0$, for a band of energy levels within order one of $\Delta=c/6$.\footnote{
Below in \S\ref{sec:3d-friendly} will spell out the role of this band of split levels around $\Delta = c/6$.}

At that point, we may start our $\eta=-1$ segment of the trajectory, with energy levels of the form \eqref{eqn:S1-En-gen}, matching the $T\bar T$-dressed $\Delta \simeq c/6$ level at $y=y_0$ with an energy level that similarly has vanishing square root.  

Consider again the energy formula \eqref{eq:Eold} for $\eta=-1$ and $\Delta=0$.  This is the Brown-York energy of a cutoff region of the static patch not containing the cosmic horizon (the pole patch in the right hand picture in figure \ref{fig:3dfriendly}).    
We can immediately write down the energy formula for a $T\bar T+\Lambda_2$ trajectory which builds up the patch complementary to this region, which we will call the cosmic horizon (CH) patch.  It is given simply by flipping the sign of the square root:
\begin{equation}\label{eq:match-trajectories-HP-level}
        {\cal E}=\frac{L^2}{\pi\lambda} \left(1 {\textbf{+}} \sqrt{-1 - \frac{4\pi^2\lambda}{L^2}\left(- \frac{c}{12}\right)}\right) = \frac{1}{\pi y} \left(1 {\textbf{+}} \sqrt{-1 + \frac{y}{y_0}}\right)\,
\end{equation}
since this corresponds to flipping the sign of the extrinsic curvature, thus capturing the complementary fill-in of the bulk region starting from the Dirichlet wall cutoff surface.  
We see that also for the cosmic horizon patch when $y=\lambda/L^2$ reaches $y_0$ \eqref{eq:y0}, the square root vanishes.  

Putting all this together
we may prescribe a combined trajectory as follows.  First, we do a $T\bar T$ trajectory as in \eqref{eq:Eold}, increasing $y$ from $y=0$ until $y$ reaches $y_0$ \eqref{eq:y0}.  
Then we evolve with $T\bar T+\Lambda_2$ increasing $y$ from its initial value of $y_0$ and choosing the branch of the square root indicated in \eqref{eq:match-trajectories-HP-level}.

As advertised, this smoothly dresses the 
$\Delta \simeq c/6$ BTZ black hole to the dS vacuum.  We can identify the microstates comprising the $dS_3$ observer patch as given explicitly by the dressed $\Delta \simeq c/6$ states.  The Cardy entropy of these states, matching the horizon entropy $2\pi\ell/4 G_N$, carries through the entire deformation since this pertains energy level by energy level.\footnote{This is sometimes described as an integrable deformation, though we stress that the seed theory need not be an integrable QFT (and our seed holographic CFTs of interest are not integrable).}

 Moreover, the agreement extends at least to the first logarithmic correction \eqref{eq:Scorrected} in the entropy \cite{Anninos:2020hfj} at the level of pure gravity, as already derived in \cite{Shyam:2021ciy}.  This follows immediately because the same result $S=A/4G_N-3\log(A/4G_N)$ arises in the AdS case as computed by Sen and Carlip \cite{Sen:2012dw}, \cite{Carlip_2000}. This density of states of the $\Delta \simeq c/6$ band of energy levels goes along for the ride in the $T\bar T$ and $T\bar T+\Lambda_2$ trajectories, for the pure gravity subsector of the theory.  It would be very interesting to extend this to all orders as in \cite{Anninos:2020hfj}\ (for a pure gravitational subsector of the theory) and explore the role of edge mode degrees of freedom as in \cite{Anninos:2021ihe}. 

Here we focused on the dressing of a particular band of energy levels near $\Delta=c/6$.  At the level of the $T\bar T+\dots $ deformed CFT, there are multiple ways to extend this readily to account for other energy levels.  For the purpose of $dS_3$ holography, we will describe the relevant prescription next in \S\ref{sec:3d-friendly}. In this context, it is the energy level we already described which contains the information about the theory that is accessible at the level of the pure gravity prescription we have formulated.  In our prescription in \S\ref{sec:3d-friendly}, the higher energy levels will not remain in the real spectrum of the theory; the total dimension of the real Hilbert space captures the $dS_3$ Gibbons Hawking entropy (in addition to a log correction \cite{Anninos:2020hfj, Shyam:2021ciy}).  In order to capture the physics of the model-dependent energy levels with $\Delta < c/6$, which are generically subleading at large $c$, we generally require additions to the specifications of the trajectory (as laid out in some cases in \cite{Hartman:2018tkw}).  As such, the energy levels in this window $\Delta < c/6$ need not be (and are not) continuous in the transition from between $\eta=1$ and $\eta=-1$.  

As it turns out, there is another way to generalise the energy level dressing \eqref{eq:match-trajectories-HP-level} of $\Delta \simeq c/6$ to other levels in the seed, based on matching the bulk geometries for black holes with $r_h \neq \ell$ following the same steps as \eqref{eqn:bh-coord}--\eqref{eq:nh-next-wsqrd-term}. We present this in \S\ref{sec:2d-friendly}. Interestingly, levels with $\Delta > c/6$ dress to locally $dS_3$ geometries with undesirable conical excesses; but these levels are screened out with a Hagedorn transition.

 A standard result from the original $T\bar T$ studies on a two-dimensional cylinder is the super-luminal propagation of perturbations.  On the gravity side of the holographic version laid out in \cite{McGough:2016lol}, this is clear from the statement that perturbations travel more quickly in the bulk bounded by a finite cylinder (rather than along the boundary).  It is not a pathology, being consistent with causality in the gravity theory, but does complicate the analysis of causal and entanglement wedges in the theory \cite{LLST}.\footnote{Interestingly, this issue does not arise for the case where the boundary theory lives on $dS_2$ \cite{GST}, which can be similarly understood on both sides of the duality.} 
We may ask the same question for $T\bar T+\Lambda_2$ on the cylinder.  This retains the faster-in-the-bulk propagation for pole patch formulated by the original trajectory \cite{GST, LLST}.  By the same token, it is sub-luminal, faster to propagate on the boundary, for the cosmic horizon patch. 
One can show that this lines up with the behavior of the boundary theory including for entanglement-based measures of propagation speed \cite{Shaghoulian:2021cef,CSY}.

Finally, without getting into details, we can describe the Hawking-Page generalization rather simply with the aid of figure \ref{fig:3dfriendly}:  below the transition, the spatial circle size $L$ can shrink to zero in the bulk, while above the transition it is $\beta$ that can shrink to zero, producing a horizon with a finite area.  These features are evident in the distinction between the non-CH patch and the CH patch in de Sitter:  in the non-CH patch the circle shrinks to zero at the north pole, while in the CH patch we get a horizon (where $\beta$ shrinks to zero in the interior but $L$ stays finite, taking the value $2\pi \ell$).  Finally, we note that one may formulate a thermofield-double system as in figure \ref{fig:ds-phases} and join the non-CH and CH patches together, integrating over their identified boundaries, to obtain a formulation of global $dS_3$ (consistent with its full symmetry group). 
We will comment on this further below, see figure \ref{fig:ds-phases}.

\section{Further details of $dS_3$ holography and microstates from $T\bar T+\Lambda_2$ }\label{sec:3d-friendly}

{
Recall first that 
ref. \cite{Dong:2018cuv} showed that the Cardy entropy of the $\Delta = c/6$ energy level matches the dS Gibbons-Hawking entropy (understood as an entanglement entropy between the two sectors of the dS/dS correspondence) if we identify the neck of this BTZ black hole with the dS neck.  This extends universally to higher dimensions, where the agreement works for AdS black holes at the Hawking-Page transition. The way that the $\sinh^2(w/\ell)$ warp factor of AdS/dS morphs into the $\sin^2(w/\ell)$ warp factor of dS/dS \cite{Alishahiha:2004md} was clarified  in \cite{GST, LLST}, by showing that the $\Lambda_2$ addition to $T\bar T$ gives a holographic description of the bulk dS rather than bulk AdS geometry in various patches. This used the prescription reviewed above in Sec. \ref{subsec:previous}, joining the trajectories at $\lambda/L^2 \gg 1$.  But the count of microstates here has remained obscure given the fact that the vacuum is dressed to the vacuum, and the $\Delta=c/6$ energy level (which contains the GH number of states) becomes complex in the dressed energy formula. Furthermore, the remaining step of joining the two dS/dS sectors by path-integrating over the metric at the central slice was not taken in \cite{GST, LLST}.        

In this section we present more details of a prescription which  provides a more direct count of microstates for the dS entropy as described above.  Instead of joining the $T\bar T$ and $T\bar T+\Lambda_2$ trajectories at $\lambda/L^2\to\infty$, we join them at a horizon --- the $\Delta=c/6$ (HP) BTZ horizon to the dS static patch horizon. That is, we bring in the boundary on the AdS side, stopping at the value of $\lambda/L^2$ for which the Dirichlet wall skirts the horizon of the HP black hole. The patch captured by the near horizon region of the BTZ HP black hole is indistinguishable from the near-horizon region of the de Sitter static patch. We start the $T\bar T+\Lambda_2$ deformation from this value of $\lambda/L^2$, and move the boundary back out, obtaining instead the dS static patch.  This deformation uses the opposite sign of the square root compared to \cite{GST, LLST}, since the bulk patch we formulate is on the other side of the boundary.  
The full static patch is obtained as a region bounded by a small cylinder at the pole (the region outside the cylinder).

\subsection{Energy Levels Beyond $\Delta \simeq c/6$}

If we move off along the $T\bar T+\Lambda_2$ trajectory starting at some nonzero value $y_0$, the dressed energy levels now take the form
\be\label{yzerodressed}
{\cal E}=\frac{1}{\pi y}\left(1\mp\sqrt{\eta+\frac{y}{y_0}(1-\eta)-4\pi^2 y\left(\Delta-\frac{c}{12}\right)+4\pi^4 y^2J^2} \right)\,,
\ee
{for states accessible at the level of pure gravity,}
where we have fixed the integration constant $C_1$ of (\ref{eqn:S1-En-gen}) in terms of the joining point $y_0$. 
If we kept the same branch of the square root, 
this formula would be manifestly continuous for all energy levels and all angular momenta, since the $\eta$ dependence drops out at $y=y_0$.  However, the pressure component is not continuous in this case.

We want to require continuity of the dressed energy for the entropically dominating HP level $\Delta=c/6$, $J=0$. Also, since the Cardy count comes from an $O(1)$ window \cite{Mukhametzhanov:2019pzy}
\be\label{eq:window}
\frac{c}{6}< \Delta \le \frac{c}{6}+ \delta
\ee
with $\delta \sim 1$, we need to make sure that the corresponding dressed states stay real. These requirements are satisfied if the matching point corresponds to the horizon of the BTZ black hole with
\be
\Delta_* = \frac{c}{6}+\delta\,,
\ee
namely
\be\label{eq:y0hor1}
y_0= \frac{1}{4\pi^2 (\Delta_*-c/12)}\,.
\ee
The minus branch from $\eta=+1$ is matched to the plus branch for $\eta=-1$ and this results in a continuous stress-tensor across $y=y_0$ for this level.

The resulting dressed energy for $y>y_0$ and $J=0$ becomes
\be\label{eq:Em2}
\mc E_{\eta=-1} = \frac{1}{\pi y} \left(1 + \sqrt{-1+4\pi^2 y \left(\Delta_*-\frac{c}{12}+(\Delta_*-\Delta) \right)}\, \right)\,.
\ee
The upper limit $\Delta=\Delta_*$ gives a stress tensor that matches continuously by construction at $y=y_0$. For the remaining states in the window (\ref{eq:window}) there are $1/c$-suppressed discontinuities in the stress-energy.\footnote{These states translate into particle states with $O(1)$ negative masses, corresponding to small conical excesses which are degenerate with $1/c$ corrections including quantum effects.}  Therefore the Cardy entropy is not modified at leading order in $c$. States with $\Delta>\Delta_*$ have complex dressed energies at $y=y_0$; they are truncated from the spectrum. We focused here on $J=0$, but it was shown in \cite{Hartman_2014} that nontrivial spins do not change the basic structure of the spectrum and entropy count.

This achieves our goal, providing an explicit microstate count for the de Sitter entropy in terms of dressed BTZ black hole states with real dressed energies.
}

\subsection{The sparse light spectrum (subleading at large $c$)}

\begin{figure}
    \centering
    \includegraphics[width=0.5\linewidth]{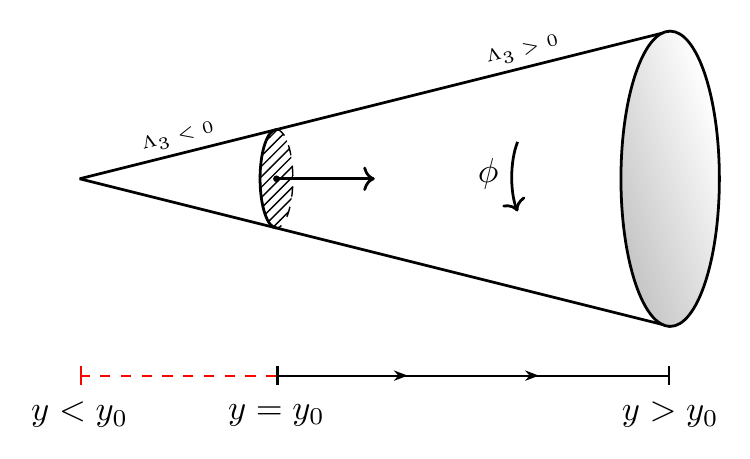}
    \caption{ \small \sffamily  A schematic of a time slice of the bulk dual for dressed $\Delta < c/6$ states. The spatial geometry piecewise forms a cone. A domain wall sits between bulk regions with different signs of $\Lambda_3$. Deforming with $T\bar{T}+\Lambda_2$ to $y>y_0$ has the effect of increasing the volume of the positive-$\Lambda_3$ domain. The radius of the internal circle decreases away from the wall at $y=y_{0}$ on both sides, unlike in the diagram.}
    \label{fig:Lambda3DomainWallCone}
\end{figure}

Consider for completeness now the dressed $\Delta<c/6$ states.  These are sparse in the holographic seed CFT and correspond to model-dependent particle states in the bulk theory.  In the $T\bar T$ part of the trajectory, when we deform until $y=y_0$ \eqref{eq:y0}\ we reach the horizon of the $\Delta=c/6$ black hole states.   As such, in the case of the $\Delta < c/6$ states we do not reach the particle position  before moving out with $T\bar T+\Lambda_2$.  Thus on the gravity side, these states entail a domain wall between bulk regions with $\Lambda_3<0$ and $\Lambda_3>0$. { We very schematically depict a time-slice of this model-dependent state in Figure~\ref{fig:Lambda3DomainWallCone}.} However, in order to capture that accurately, one would require additions to the trajectory going beyond pure gravity (building from the results in \cite{Hartman:2018tkw}).  

\subsection{Lazarus States: $\Delta > c/6$ at large $y$}

An interesting note regarding the formula \eqref{yzerodressed} is that the energies are real for
\begin{equation}
    \Delta < \Delta_{max} \approx \frac{c}{4} - \frac{1}{4\pi^2 y}.
    \label{eqn:3d-delta-max}
\end{equation}
For $y = y_0$, the RHS is $c/6$, as required by consistency with the previous discussion. But the value of $\Delta_{max}$ \emph{increases} with $y$.

This means that as we increase $y$ to $y_0$, flip $\eta$ and continue along the trajectory, many states with $\Delta > c/6$ go complex and then become real again --- they are resurrected, as in the Biblical story of Lazarus. 
We do not currently have a proposed bulk interpretation for these states.  Of course, if we maintain unitarity throughout this process by restricting the theory to its real spectrum at every step, there is no reason to include them again.

\section{Modular Behavior And A Role for the Hawking-Page Transition}\label{sec:HP-mod}

The dressed $\Delta \simeq c/6$ band of energies accounting for the microstates has a special status in the seed CFT:  this energy level corresponds to the Hawking-Page transition \cite{Hartman:2013qma, Dong:2018cuv}. In this section, we note that the role of these microstates could be motivated starting from the original trajectory \cite{GST, LLST}\ (reviewed in \S\ref{subsec:previous}) and applying a remnant of modular invariance in our system.

 We may work in a canonical or microcanonical ensemble, with fluctuations suppressed at large $c$.  In the gravity variables, these are spelled out in the Brown-York papers \cite{Brown:1992br} and \cite{Brown:1992bq}.
Consider a partition function for a two-dimensional theory on a torus of size $L \times \beta$.
For simplicity, let us set the spin to $0$ and take the sides of the torus to be orthogonal.
The modular S-transformation takes this torus to one of dimensions $\beta \times L$.
This is a symmetry, due to the fact that there is no special choice of time on this Euclidean torus,
\begin{equation}
  Z_{\lambda} (L,\beta) = Z_{\lambda} (\beta,L).
  \label{eqn:s-invariance}
\end{equation}
This trivial property, which is just a special case of the diffeomorphism-invariance of the theory, is called S-invariance.\footnote{In some of the literature, the S-transformation is defined not just as a diffeomorphism but as a diffeomorphism followed by a rescaling $L \times 	\beta \to L \times L/\beta$, following the CFT literature. That is a symmetry if we also take $\lambda \to \lambda/(\beta/L)^{2}$. The final answers don't depend on which symmetry we work with.  See e.g. \cite{Datta:2018thy,Aharony:2018bad, Aharony:2018ics} for previous studies of modular properties of $T\bar T$ deformed theories.}
As the deformation \eqref{eqn:ttbar-eqn} is diffeomorphism-invariant, we expect it to also be S-invariant, at least before discarding the complex energy levels (at which point the theory reduces to ordinary $0+1$-dimensional quantum mechanics).  In fact, even restricting attention to the real energy levels leaves us with a residual S-invariance as follows.

In large $c$ field theories with a sparse spectrum, for $\beta$ sufficiently close to $L$, the partition function takes the form
\begin{equation}
  \log Z \simeq \max \left\{ - \beta E_{vac} (L) \ ,\ - L E_{vac} (\beta) \right\}
  \label{eqn:S-inv-Z-new}
\end{equation}
to leading order at large $c$ \cite{Hartman_2014}.  Starting from this seed CFT, the $T\bar T$ and $T\bar T+\Lambda_2$ deformations proceed energy level by energy level \cite{Smirnov:2016lqw,GST,LLST}.
Clearly, the first term dominates if $L < \beta$ and the second if $L > \beta$. 

It is interesting to consider the bulk dual corresponding to the two terms.
 To do so, we have to embed the torus into the Euclidean version of the static patch metric \eqref{eqn:static-patch} (where the Euclidean time $\tau_E$ is constrained to have a periodicity of $2\pi \ell$ by smoothness).
From the bulk metric, an $r = r_c$ surface is a torus of dimensions $2\pi r_c \times 2\pi \sqrt{\ell^2 - r_c^2}$.
By the matching conditions \eqref{eq:param-dictionary}, this constrains the boundary modular parameter to be
\begin{equation}
    \left( \beta/L \right)^2 = \frac{y}{y_0} - 1.
    \label{eqn:beta-y-reln}
\end{equation}
So we see that, unlike a standard two-dimensional theory, the modular parameter and the coupling cannot be varied independently.\footnote{This is the case if we restrict to global $dS_3$. More generally, we may allow conical deficits, in which case \eqref{eqn:beta-y-reln} becomes an inequality --- owing to the fact that conical excesses are not allowed. This is still surprising from the two-dimensional point of view.} 

In the recent work \cite{Shyam:2021ciy}, the second term of \eqref{eqn:S-inv-Z-new} in the $T\bar T+\Lambda_2$ deformation, in the limit $\beta/L \to 0$, accounts for the leading (and first subleading) contribution to the Gibbons-Hawking entropy.  
So, let us restrict attention to the second term, where the contractible cycle in the bulk is identified with the time cycle in the boundary.
Assuming then that $\beta < L$, we can write \eqref{eqn:S-inv-Z-new} as
\begin{equation}\label{eq:Z-with-c6}
    \log Z|_{\beta < L} \simeq  - L E_{vac} (\beta) = S_{Cardy} (\Delta = c/6) - \beta E_{\Delta = c/6} (L)
\end{equation}
with $E_{vac}$ given by the dressed energy (\ref{eq:Eold}) with $\Delta=0$, and $E_{\Delta=c/6}$ the $T\bar T+\Lambda_2$ dressed energy \eqref{yzerodressed} corresponding to the cosmic horizon patch. 
The crucial second equality  follows from \eqref{eqn:beta-y-reln}. A proof of \eqref{eq:Z-with-c6} is explained in more detail in Appendix \ref{app:vac-and-c6}. 

Note that $E_{vac}$ is the vacuum energy in the trajectory of \cite{LLST} (discussed in \S\ref{subsec:previous}), whereas $E_{\Delta=c/6}$ is an energy in the new trajectory defined in \S\ref{subsec:new}.
In \S\ref{sec:review}, these two trajectories were introduced merely as two versions of the $T \bar{T} + \Lambda_2$ theory, but now we see that the partition function somehow puts them in the same theory, as two different sectors.
The vacuum state in the old \S\ref{subsec:previous} trajectory seems to be mapped to the $\Delta=c/6$ state in the new \S\ref{subsec:new} trajectory.
To be clear, in this section we have not allowed arbitrary energies to appear --- thanks to the constraint \eqref{eqn:beta-y-reln}. In that sense, we have discussed a one-dimensional theory which has only certain states and cannot have full modular S-invariance.
However, because of the large $c$ thermodynamic limit, the partition function in the CH phase is dominated by states with $\Delta \simeq c/6$; thus it is reasonable for the vacuum to map to only this window of states in some remnant of S-invariance that shows up in a partition function constrained by \eqref{eqn:beta-y-reln}.
Thus, demanding that the theory defined by the energy levels in \cite{LLST} satisfy this constrained version of S-invariance automatically implies that the theory also contains energy levels corresponding to the Gibbons-Hawking entropy.

This is analogous to the Hawking-Page transition in AdS spacetime. For the Cosmic Horizon patch, considered in Euclidean signature, the thermal circle shrinks smoothly in the interior (at the horizon).  The complementary Pole Patch has the spatial circle shrinking smoothly at the pole.  
On the gravity side, we have defined a Dirichlet boundary condition that the boundary be a torus of given dimensions and the path integral is a sum over all 3d manifolds that satisfy this boundary condition.
So, we find
\begin{align}
    \log Z = - \min ( I_P\ ,\ I_{CH} ) = - \frac{\beta L}{\pi \lambda} \min \left\{ 1 - \frac{\beta}{L}\ ,\ 1 - \frac{L}{\beta} \right\}
    \label{eqn:3d-Z}
\end{align}
where $I_P$ and $I_{CH}$ are the classical Euclidean actions for the Pole and Cosmic Horizon patches. We supply the details of this calculation in Appendix \ref{app:grav-actions}.  
We see explictly that the pole (CH) saddle dominates when $\beta > L$ ($\beta < L$).

In the way that the two saddles correspond to a choice of contractible boundary cycle, the role of the two terms here is the same as the standard Hawking-Page transition in AdS/CFT. 
However, there is an important difference: in the present case we don't have a tower of energy levels, as shown by our energy formula.
To see that this is consistent with the formulas above, we merely have to take the $\beta \to 0$ limit of \eqref{eqn:3d-Z} to find the dimension of the Hilbert space \cite{Caputa:2020fbc}.
We find that the action approaches $1/(\pi \sqrt{y_0 y}) = \sqrt{c/(3y)}$.

\begin{figure}[t!]
  \centering
  \includegraphics[width=0.425\linewidth]{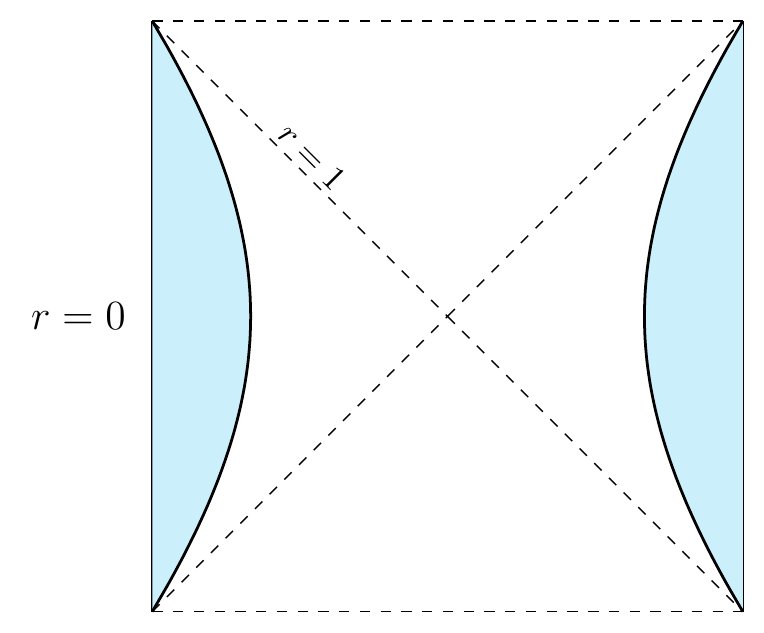}\hspace{0.1\linewidth}
  \includegraphics[width=0.425\linewidth]{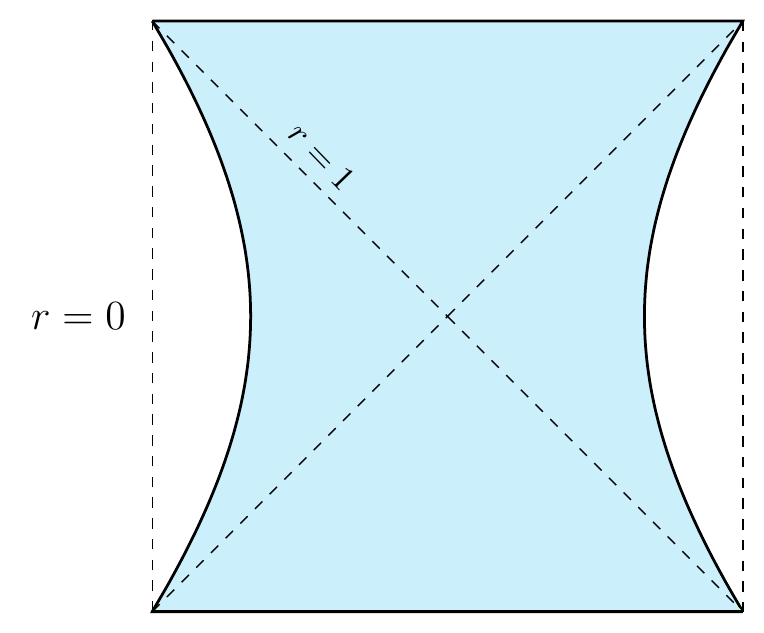}
  \caption{\small \sffamily  The two phases of the doubled system, including the entangled cosmic horizon patches in the right panel. In that case, the upper and lower triangles should be reconstructable by bulk evolution.  One may treat these patches as conditioned on an observer at the Dirichlet wall boundary, or alternatively use them as building blocks, joining these patches together at their common boundary, integrating over its geometry, to obtain global $dS_3$ with its symmetries. It would be interesting to connect the full nonlinearity of the deformed theory as a function of the fundamental seed CFT variables to the notions of hyper-scrambling and complexity suggested recently in \cite{Susskind:2021esx,Chapman:2021eyy}.}
  \label{fig:ds-phases}
\end{figure}

Finally, let us note that we may formulate a thermofield double state of two copies of our system, 
\begin{equation}\label{eq:TFD-gen}
    |TFD \rangle = \sum_n e^{-\beta E_n/2}|E_{n,L}\rangle |E_{n,R}\rangle
\end{equation}
whose norm is $Z(\beta)$.  
The corresponding bulk dual is as in figure \ref{fig:ds-phases}.
Note however that, because of \eqref{eqn:beta-y-reln}, we cannot describe the CH patch with a boundary arbitrarily close to the poles  in the canonical ensemble.  This patch was, however, formulated above in \S\ref{sec:review}-\ref{sec:3d-friendly} in the microcanonical ensemble, leading to our microstate count for arbitrary $y\in (y_0, \infty)$.  In the canonical ensemble, on the other hand,
since this saddle only dominates when $\beta < L$, we find that the boundary of the CH patch must satisfy
\begin{equation}
    y < y_{max,CH} = 2 y_0 = \frac{6}{c \pi^2} \qquad \Rightarrow \qquad r_c^2 > \frac{\ell^2}{2}.
    \label{eqn:phase-bd}
\end{equation}
Interestingly, since this is the limit $\beta = L$, the boundary of the pole patch must satisfy $r_c^2 < \ell^2/2$  for the dominant contribution in the canonical ensemble.

\section{An Alternative Deformation with a Hagedorn Transition}\label{sec:2d-friendly}

In this section, we will explore another version of the $T\bar T+\Lambda_2$ deformed CFT which again makes use of the freedom to choose the integration constant and branch of the square root in the energy formula \eqref{eqn:S1-En-gen}.  This version will exhibit a Hagedorn transition at the would-be Hawking-Page transition point $\beta=L$, leading to different physics above that scale.  This version agrees with the above prescription in \S\ref{sec:3d-friendly} for the microstates of interest --- at least in the near-horizon regime ---, but rather than turning the higher levels complex (so dropped in a unitary theory), it prescribes a high-energy spectrum that leads to a Hagedorn divergence in the naive partition function above the dressed $\Delta=c/6$ energy level.  It may be that this Hagedorn behavior similarly screens the system from unphysical higher energy levels --- indeed in the fundamental string version of a Hagedorn transition, there is a winding string condensate that has been argued to reduce the number of degrees of freedom in the system \cite{Atick:1988si}.   In this work we will not pursue this transition further, returning our attention to the microstates captured in the prescription of \S\ref{sec:review}-\ref{sec:3d-friendly}.        
The simplest way to think about this new theory is to imagine following the steps of the matching in \eqref{eqn:bh-coord}--\eqref{eq:nh-next-wsqrd-term} that related the $dS_3$ vacuum to the $r_h = \ell$ BTZ black hole, but for a black hole with $r_h > \ell$. The only role that the precise value of $r_h$ played in the matching above was in ensuring that on the $dS$ side the horizon had the right area --- which is fixed by the global smoothness of the solution, So, the cost we pay for increasing $r_h$ on the BTZ side is to introduce a pair of conical \emph{excesses} at the poles of $dS$; the Penrose diagram remains identical to that in Fig.~\ref{fig:3dfriendly}.

We find for the energies
\begin{equation}\label{eq:2d-energies-LLSTconv}
    \mathcal{E} = \frac{1}{\pi y}\left\{ 1 + \sqrt{-1 + 4\pi^2 y \left(\Delta - \frac{c}{12} \right) + 4\pi^4 J^2 y^2 } \right\} \,.
\end{equation}
In terms of a trajectory, this corresponds to flipping the value of $\eta$ at a state-dependent value
\be
y_0= \frac{1}{4\pi^2(\Delta-c/12)}\,.
\ee
The bulk description of these states in the case $J=0$ is the patch 
\begin{align} \label{eqn:ch-metric}
    ds^2 = -\left( 1 - \frac{\rho^2}{\ell^2} \right) d\tau^2 + \frac{d\rho^2}{1-\frac{\rho^2}{\ell^2}} + \rho^2 d\phi^2, \qquad\qquad  \rho \in \left[ \rho_c ,\ell \right],\, \phi\sim \phi + 2\pi \mu,
\end{align}
where
    \begin{align}
        \rho_c = \frac{\ell}{\mu}\sqrt{\frac{3}{c\pi^2 y}}=\frac{L}{2\pi\mu}, \qquad\qquad  \mu = \frac{1}{\sqrt{(4\pi^2 y)(c/12) - (\beta/L)^2}}.
    \end{align}
The parameter $\mu$, the weight of the conical defect, can be fixed by calculating the Brown-York stress tensor, giving
\begin{equation}
    \Delta - \frac{c}{12} = \frac{c}{12} \mu^2 \qquad\qquad \left(\Delta > \frac{c}{12}\right).
\end{equation}
For $\Delta>c/6$, this would correspond to a conical {\it excess} sourced at $\rho = 0$.  By increasing $\Delta$, this would formally give an arbitrarily {\it negative} mass source at this point in the bulk. The reason this is not an out-and-out inconsistency is that these negative-mass sources are outside the patch described by our putative boundary theory. Regardless, since our interest is a theory of $dS_3$ we would hope that they are excluded in some fashion.

As it happens, this may very well be the case.
To see this, let us ask whether there is any reason to consider this alternative theory at all, given the double strangeness of a state-dependent $y_0$\footnote{We should clarify that this is an aesthetic problem; a theory with the above energy levels is a perfectly valid solution to the flow equation. It is also connected to the seed CFT in a
novel and interesting way.} and conical excesses.
The answer is that these conical excesses naturally arise when trying to extend the considerations of \S\ref{sec:HP-mod} beyond boundary modular parameters satisfying the global $dS_3$ constraint \eqref{eqn:beta-y-reln}.\footnote{The weight of the conical singularity in the canonical ensemble is
\begin{equation*}
    \mu^2 = \frac{1}{(y/y_0) - (\beta/L)^2}.
\end{equation*}
}
Happily, studying this partition function further reveals the existence of a Hagedorn transition that intervenes before these naive (and problematic) states enter.  

Indeed, let us check the Hagedorn temperature for the naive spectrum \eqref{eq:2d-energies-LLSTconv}.  The partition function takes the form\footnote{The density of states is the Cardy density of states in terms of the seed dimension $\Delta$; this can be checked by an inverse Laplace transform of $I_{CH}$ on a background that includes these excesses.}
\begin{equation}\label{eq:Z-as-sum}
    Z(\beta, L)= \sum_{\Delta} \rho(\Delta) \exp(-\beta {\cal E}/L)
\end{equation}
where in this expression $\rho(\Delta)$ is the number of states at the energy level  $\frac{2\pi}{L}(\Delta-\frac{c}{12})$ in the seed theory (this number stays constant throughout the deformation at the level of pure gravity).
In order to locate the Hagedorn transition temperature, we must check convergence of this expression.  For fixed deformation parameter $y$ and a given $\beta$, the question of convergence is determined by the behavior of the summand (or integrand in a continuum treatment) at large $\Delta$, where the density of states take the Cardy form (as it does for all levels starting at $\Delta=c/6)$, and where the $\Delta$ term inside the square root dominates in the energies \eqref{eq:2d-energies-LLSTconv}.  So the summand/integrand boils down to
\begin{equation}\label{eq:Zintegrand}
    \exp\left(2\pi\sqrt{\frac{c}{3}\left(\Delta-\frac{c}{12}\right)}-2\frac{\beta}{L}\sqrt{\frac{\Delta -\frac{c}{12}}{y}}\right)
\end{equation}
meaning that there is a Hagedorn transition,
\begin{equation}\label{eq:Hagedorn}
    Z < \infty \qquad \text{when} \qquad \left. \frac{\beta}{L} > \frac{\beta}{L}\right|_{Hagedorn}=\pi\sqrt{\frac{c y}{3}} = \sqrt{\frac{y}{y_0}}.
\end{equation}

Since we don't have the constraint \eqref{eqn:beta-y-reln} in this case, this system exhibits a Hawking-Page transition at
\begin{equation}\label{eq:HP}
    I_{CH} < I_{P} \qquad \text{when} \qquad \left. \frac{\beta}{L}\right|_{Hawking-Page}<1.
\end{equation}
Since we want to describe the CH patch, we need to check that the inequality \eqref{eq:Hagedorn} is consistent with this one.
Clearly, the answer to this question depends on $y$. 
In the trajectory defined here, $y$ ranges from $y_0$ where we switch to the $T\bar T + \Lambda_2$ part of the trajectory, to some $y_{final}$ which we will take to be state-independent.  

First, note that if we took $y_{final}\to\infty$ (to capture the full static patch with identifications) it would lead to a vanishing Hagedorn temperature.
To avoid this, let us consider a $y_{final}$ which is common to all levels, and as small as possible given that. This is a $y_f$ which is equal to the $y_0$ value for the $\Delta = c/6$ level,  
\begin{equation}
    y_{final}\equiv y_{0} = \frac{1}{4\pi^2(\frac{c}{6} - \frac{c}{12})} = \frac{3}{c\pi^2}
\end{equation}
and we find
\begin{equation}\label{eq:Hag-for-yfc6}
        \left. \frac{\beta}{L}\right|_{Hagedorn}=\pi\sqrt{\frac{c~y_{final}}{3}}= \pi\sqrt{\frac{c~ y_{0}}{3}}=1= \left. \frac{\beta}{L}\right|_{HP}\,.
\end{equation}
As a result, the partition function of this version of a $2d$ $T\bar T+\Lambda_2$ trajectory simply does not converge above the HP=Hagedorn transition, indicating a phase transition.\footnote{Indeed, this value of $y_{final}$ is minimal:  
from this analysis, the partition function converges for
$\frac{\beta}{L}\ge \pi \sqrt{\frac{c \lambda}{3L^2}}$.
Since we used S-invariance in the approach of this section, we also need to demand this inequality under the exchange $\beta \leftrightarrow L$, $\frac{L}{\beta}\ge \pi \sqrt{\frac{c \lambda}{3\beta^2}}$.
This last inequality requires $\frac{c\,\pi^2}{3} y \le 1\;\Rightarrow\; y\le y_0$}
This phase transition is complicated in the  analogous case of fundamental string theory \cite{Atick:1988si}, but it may provide a self-censorship of the problematic high energy spectrum \eqref{eq:2d-energies-LLSTconv}.  

Further, since we are interested in describing the CH patch with no conical excesses, we must compare with the global $dS_3$ condition \eqref{eqn:beta-y-reln}.
Clearly,
\begin{equation}
    \left. \frac{\beta}{L} \right|_{Hagedorn} = \sqrt{\frac{y}{y_0}} > \left. \frac{\beta}{L} \right|_{global} = \sqrt{\frac{y}{y_0} - 1},
\end{equation}
and so the partition function doesn't converge in the requisite regime.
It is of course possible that the scale of new physics is at high enough temperature that the global $dS_3$ regime isn't screened, but we cannot check this with the tools at our disposal.

Thus, we are led to focus on the prescription in \S\ref{sec:review}-\ref{sec:3d-friendly}\ accounting for the microstates as dressed $\Delta=c/6$ levels (related to modular transformations and the Hawking-Page transition as summarized in \S\ref{sec:HP-mod}).

\section{Discussion}

In this work, we exhibited a concrete set of microstates accounting for the Gibbons-Hawking entropy $S_{GH}$.  The  microstates, a band of energy levels obtained as dressed $\Delta\simeq c/6$ states in the seed CFT, are accessible at the level of pure gravity, which is formulated by the $T\bar T+\Lambda_2$ deformation.  This theory also reconstructs the de Sitter geometry as in \cite{GST}\cite{LLST}.  The finiteness of $S_{GH}$  fits with the finiteness of the real dressed spectrum of the theory.  
The results extend to corrections such as those treated in \cite{Anninos:2020hfj, Shyam:2021ciy} as a dressing of the corresponding AdS/CFT results in \cite{Sen:2012dw}; it will be interesting to connect also to recent studies such as \cite{Anninos:2021ihe},\cite{hikida2021holography} identifying microphysical degrees of freedom from another approach, as well as top-down methods such as \cite{Dong:2010pm}.

The theoretical control we have in this work is based on the universality and solvability of the $T\bar T+\Lambda_2$ theory, in the same sense as in the original $T\bar T$ deformation \cite{Zamolodchikov:2004ce, Smirnov:2016lqw, Dubovsky:2018bmo}.  This universal and solvable theory captures the physics that is available in the pure gravity sector.  A full accounting of finer structure requires additional generalizations, as introduced e.g. for bulk scalar fields in \cite{Hartman:2018tkw}.  Even at the level of pure $T\bar T$ and $T\bar T+\Lambda_2$, although calculable in terms of the dressed energy spectrum, these theories remain somewhat mysterious.   We may regard the holographic matching of energies and entropies -- even in the case of bulk de Sitter spacetime -- as ample motivation to improve our understanding of this space of deformed theories.  It explicitly captures essential features of the physics, but it may have limitations in its regime of validity; we comment on this further below in the context of late-time physics in cosmological quantum gravity.  

\subsection{Remarks and future directions}

Finally, in the remainder of this section we will comment on additional aspects of our theory and future directions.  

\subsubsection{Other dimensionalities}

We have exhibited the Gibbons-Hawking microstates specifically for a $dS_3$ bulk theory in the bulk of this paper.  At least at large $N$, similar methods apply in other dimensionalities: the $\Lambda_2$ deformation generalizes to a similar $\Lambda_d$ contribution to the appropriate analogue of the $T\bar T$ deformation, providing a building block for formulating $dS_{d+1}$.  Analogues of the $T\bar T$ deformation in other dimensionalities have been treated in a number of works including \cite{Hartman:2018tkw, Gross:2019ach, Iliesiu:2020zld, Stanford:2020qhm, Griguolo:2021wgy}.  

In all dimensionalities it was noted in \cite{Dong:2018cuv} that black holes at the Hawking-Page transition at large $c$ account for the Gibbons-Hawking entropy for de Sitter with the same horizon size; moreover at the HP level the warp factor in the $AdS_{d+1}$ geometry matches that of $dS_{d+1}$, with both foliated by $dS_d$ slices.  The static patch Hamiltonian is the modular Hamiltonian for the dS/dS correspondence \cite{Alishahiha:2004md, GST, LLST}, so the calculations are closely related.    It would be interesting to analyze this further.\footnote{including the question of the fate of modular invariance properties -- which played a role at intermediate steps in our analysis -- in higher dimensions \cite{Shaghoulian:2016gol}}
It is particularly important to study the case of $dS_4$ and incorporate the physics of black holes in de Sitter, a topic studied recently in e.g. \cite{Maldacena:2019cbz, Susskind:2021dfc} with an eye toward holography.   

\subsubsection{Chaos and complexity}

It was recently suggested in \cite{Susskind:2021esx} that a dual to de Sitter requires a hyper-scrambling behavior, and must generate rapidly growing complexity.\footnote{The infinite complexity of de Sitter surprising nobody who has worked on the subject.}  
 To the extent that this follows from the gravity-side calculation given there and in \cite{Chapman:2021eyy}, this was connected to the Hamiltonian needing to be non-k-local in terms of the fundamental degrees of freedom.  This means that at fixed number of degrees of freedom, the Hamiltonian must contain more than a certain monomial power of the fundamental degrees of freedom \cite{Susskind:2021esx}.   

In this regard, let us note that the $T\bar T+\Lambda_2$ deformation is irreducibly nonlinear (and hence non-k-local) in terms of the seed CFT degrees of freedom.  With $\eta=-1$ (corresponding to nonzero $\Lambda_2$), the square root formula for the dressed energies -- as well as the dressed classical action and Hamiltonian -- does not admit an expansion in the CFT variables.   (During the earlier $T\bar T$ segment of the trajectory, the Taylor expansion of the square root is available, and has radius of convergence 1, becoming highly nonlinear as we approach the matching point.)

Since the $T\bar T+\Lambda_2$ theory embodies the geometry of de Sitter space, having passed significant and detailed tests in matching the Brown-York energies and entanglement entropies \cite{GST, LLST}, it would be interesting to probe it further using these ideas about scrambling and chaos.  This approach is versatile and applies to truly large-radius gravity, e.g. for the seed CFT being the D1-D5 system, although with additional elements beyond $T\bar T+\Lambda_2$ \cite{Hartman:2018tkw} required for approximate bulk locality.   

However, in this analysis we should take into account the subtleties with arbitrarily late time physics in cosmological quantum gravity.  We will comment on that briefly in \S\ref{sec:late-times}.

\subsubsection{Joining the cosmic horizon and pole patches: global bulk $dS$ and its symmetries}

As described above, the culmination of the new trajectory we defined here is to move the boundary of the cosmic horizon patch to the pole.  This yields in the bulk the full static patch, with boundary at the observer position.  This may be viewed as a formulation conditioned on such an observer.  But it is also possible to join together the cosmic horizon and pole patches, identifying their shared boundary and path integrating over its geometry.  In the doubled system depicted in figure \ref{fig:ds-phases}, this joined system covers global $dS_3$.  As such, it {should} contain the full symmetries of global de Sitter at the level of the combined $T\bar T+\Lambda_2$ systems. Understanding the re-emergence of the full symmetry group is an important direction, at the appropriate level of approximation (given the possibility of non-perturbative decays of de Sitter spacetime).

\subsubsection{Late times}\label{sec:late-times}

There are strong indications that de Sitter eventually decays at late times, a manifest feature of string compactifications  (including those built by uplifting AdS/CFT models \cite{Dong:2010pm, DeLuca:2021pej}) since the effective potential decays to zero at weak curvature.  This is not immediately evident in the $T\bar T+\Lambda_2$ theory, which is solvable at finite $c$. But this is not a contradiction:  the details of the decay arise in the bulk scalar sector of the theory, which is model-dependent and involves the further generalization of $T\bar T$ formulated e.g. in \cite{Hartman:2018tkw}.  
Both the covariant entropy bound \cite{Bousso:1999xy} and string-theoretic
counts of microphysical degrees of freedom \cite{Dong:2011uf}\ indicate that the dimension of the Hilbert space relevant to the full decaying system increases, asymptotically approaching infinity.  It would be interesting to extend the $T\bar T+\Lambda_2$ approach to include this late-time physics.  

Even before considering this more general late-time behavior, it is interesting to consider situations where approximately de Sitter expansion does not extend arbitrarily far in the past.  For example, depending on the quantum gravitational resolution of spacelike singularities, it may happen that consistent histories exist which do not include the de Sitter neck (where its spatial sphere reaches its minimal value at the moment of time symmetry), and the system instead starts from a spacelike classically singular initial condition.  Such cases, if they are consistent in quantum gravity, may require additional states with large entropy.\footnote{Recent works exploring interesting aspects of the particular case of no boundary wavefunctions, and their implications for de Sitter entropy include \cite{Penington:2019kki} and \cite{Chen:2020tes}.}

In any case, for the present work we focused on the observer patch in $dS_3$, whose finite entropy \eqref{eq:Scorrected} precisely matches the real spectrum of our simple solvable $T\bar T+\Lambda$ deformation.  This motivates further work in many directions.

\section*{Acknowledgements}
We would like to thank Victor Gorbenko for many discussions, including stressing early on the solvability and universality of $T\bar T+\Lambda_2$ on the cylinder, and for comments on the manuscript. We would also like to thank Edgar Shaghoulian and Zhenbin Yang for useful discussions.

The work of E.S. is supported in part by a Simons Investigator award and National Science Foundation grant PHY-1720397. V.S. is supported by The Branco Weiss Fellowship - Society in Science, administered by the ETH Zürich.  G.T. is supported by CONICET, CNEA, UNCuyo and by ANPCYT PICT grant 2018-2517.  E.A.C. is supported by the US NSF Graduate Research Fellowship under Grant DGE-1656518. E.A.M is supported by a Kadanoff fellowship at the Kadanoff Center of the University of Chicago, and would also like to acknowledge the support of the Jones Endowment for Physics Research. RMS is supported by STFC consolidated grant ST/T000694/1, the Isaac Newton Trust Grant ``Quantum Cosmology and Emergent Time'' and the AFOSR grant “Tensor Networks and Holographic Spacetime.”

\appendix

\section{Conventions}\label{sec:notationconventions}

We consider the following conventions, in Euclidean signature (euclidean signs as in \cite{Kraus:2018xrn}):
\bea
T_{\mu\nu}&=& \frac{2}{\sqrt{g}} \frac{\delta}{\delta g^{\mu\nu}}\, \log Z \nonumber\\
T\bar T &\equiv& \frac{1}{8}(T_{ab}T^{ab}-(T^a_a)^2) \nonumber\\
\frac{\partial}{\partial \lambda}\, \log Z& =&-2\pi \,\int d^2 x\,\sqrt{g}\, \langle T \bar T \rangle+\frac{1-\eta}{2\pi \lambda^2}\int d^2 x \,\sqrt{g}\nonumber\\
T^\mu_\mu& =& -\frac{c}{24\pi} R^{(2)} - 4\pi \lambda  T \bar T  +\frac{1}{\pi \lambda}(1-\eta)\nonumber\\
\mc E &=& E L= (-T^\tau_\tau) L^2= \frac{L^2}{\pi \lambda}\left(1 \mp \sqrt{\eta-4\pi^2 \frac{\lambda}{L^2} \left(\Delta-\frac{c}{12} \right)+4 \pi^4 J^2 \frac{\lambda^2}{L^4} }\, \right) \nonumber\\
\Delta&=& h+ \tilde h\;,\;J= h - \tilde h\,.
\eea
Here $L=2\pi r$ is the length of the circle, the energy density $\rho= -T^\tau_\tau$, and $\lambda$ has energy dimension $-2$. We will often employ the dimensionless combination
\be
y =\frac{\lambda}{L^2}\,,
\ee
which transforms under modular transformations. In particular, the S-transformation having $L$ explicit is just $\beta \leftrightarrow L$, where $\beta$ is the length of the thermal circle; so $y \to \frac{L^2}{\beta^2} y$.

In Lorentzian signature $(-, +)$ this is:
\bea
T_{\mu\nu}&=& -\frac{2}{\sqrt{-g}} \frac{\delta}{\delta g^{\mu\nu}}\, \log S_{eff} \nonumber\\
T\bar T &\equiv& \frac{1}{8}(T_{ab}T^{ab}-(T^a_a)^2) \nonumber\\
-\frac{\partial}{\partial \lambda}\, S_{eff}& =&-2\pi \,\int d^2 x\,\sqrt{-g}\, \langle T \bar T \rangle+\frac{1-\eta}{2\pi \lambda^2}\int d^2 x \,\sqrt{-g}\nonumber\\
T^\mu_\mu& =& -\frac{c}{24\pi} R^{(2)} - 4\pi \lambda  T \bar T  +\frac{1}{\pi \lambda}(1-\eta)\nonumber\\
\mc E &=& E L= (+T^\tau_\tau) L^2= \frac{L^2}{\pi \lambda}\left(1 \mp \sqrt{\eta-4\pi^2 \frac{\lambda}{L^2} \left(\Delta-\frac{c}{12} \right)+4 \pi^4 J^2 \frac{\lambda^2}{L^4} }\, \right) \nonumber\\
\Delta&=& h+ \tilde h\;,\;J= h - \tilde h\,.
\eea
where again $L=2\pi r$ is the length of the circle, $\lambda$ has energy dimension $-2$, and we will define $y=\lambda/L^2$.   The energy density in Lorentzian signature is now $\rho= +T^\tau_\tau$.

The energies in the seed CFT are
\begin{equation}
    {\cal E}_0=E_0L = {2\pi} (\Delta -\frac{c}{12})\,.
\end{equation}

\section{Euclidean actions for the Pole patch and the Cosmic Horizon patch:  exchange of dominance at $\beta=L$}
\label{app:grav-actions}
In this appendix, we compute the Euclidean gravitational actions for the pole patch and the Cosmic Horizon patch, and show that the two saddles switch dominance at $\beta=L$.

The Euclidean version of the action \eqref{eqn:lorentz-grav} is
\begin{equation} \label{eqn:eucl-grav}
    I_E = -\frac{1}{16\pi G_N} \int_{\mathcal{M}_3} d^3 x\sqrt{g} \left(R + \frac{2}{\ell^2}\eta \right) - \frac{1}{8\pi G_N} \int_{\partial \mathcal{M}_3} d^2 x \sqrt{g} \left( K - \frac{1}{\ell} \right).
\end{equation}
The Pole patch metric with $\tau_E=i\tau$ is
\[
  ds_{3}^{2} = \left(\frac{\ell^2 - r^{2} }{{\ell^2}}\right) d\tau_E^{2} + \left(\frac{{\ell^2} }{\ell^2-r^{2}}\right)dr^{2} + r^{2} d\phi^2, 
\]
with
\begin{equation}
    r \in \left[0,r_c\right], \quad \tau_E \sim \tau_E + \beta_c,\quad \phi \sim \phi + 2\pi
\end{equation}
where
\begin{equation} \label{eqn:static-rc}
    r_c= \ell\sqrt{\frac{3}{c\pi^2  y}} = \frac{L}{2\pi},\quad \beta_c = \frac{\beta}{\sqrt{1-\frac{3}{c\pi^2 y}}} =  \frac{\beta}{\sqrt{1-\frac{r_c^2}{\ell^2}}}.
\end{equation}
From the near-horizon behavior derived above in \eqref{eq:nh-next-wsqrd-term}\ we see that $\beta_c=2\pi\ell$.

The extrinsic curvature at the boundary $r=r_c$ is
\begin{align}
    K_{{\tau_E}{\tau_E}} = -\frac{r_c}{\ell^2}&\sqrt{1-\frac{r_c^2}{\ell^2}},\quad  K_{\phi\phi} = r_c \sqrt{1-\frac{r_c^2}{\ell^2}},\quad K =  \frac{1-\frac{2r_c^2}{\ell^2}}{r_c\sqrt{1-\frac{r_c^2}{\ell^2}}} \label{eqn:pole-ext-curv}\\
    &\sqrt{g}\left(K - \frac{1}{\ell}\right) = 1- \frac{2r_c^2}{\ell^2} - \frac{r_c}{\ell}\sqrt{1-\frac{r_c^2}{\ell^2}}\,.
\end{align}
The first term in \eqref{eqn:eucl-grav} is
\begin{equation}
    -\frac{1}{16\pi G_N}\int_0^{\beta_c}d\tau_E \int_0^{2\pi} d\phi  \int_0^{r_c}dr \,\frac{4 r}{\ell^2} = -\frac{\beta_c}{4 G_N}\frac{r_c^2}{\ell^2}
\end{equation}
and the second term in \eqref{eqn:eucl-grav} is
\begin{equation}
    -\frac{1}{8\pi G_N}\int_0^{\beta_c} d\tau_E \int_0^{2\pi} d\phi \sqrt{g}\left(K - \frac{1}{\ell}\right) = -\frac{\beta_c }{4G_N}\left(1- \frac{2r_c^2}{\ell^2} - \frac{r_c}{\ell}\sqrt{1-\frac{r_c^2}{\ell^2}} \right).
\end{equation}
Adding the two terms, we get
\begin{equation}
    I_{Pole} = \frac{1}{4G_N} \frac{\beta}{\sqrt{1-\frac{r_c^2}{\ell^2}}} \left( \frac{r_c}{\ell} \sqrt{1-\frac{r_c^2}{\ell^2}}-\left(1-\frac{r_c^2}{\ell^2}\right) \right) = \frac{\beta r_c}{4G_N\ell} \left( 1-\sqrt{\frac{\ell^2}{r_c^2}-1}\right).
\end{equation}
Using the dictionary \eqref{eq:param-dictionary} and \eqref{eqn:static-rc}, we get
\begin{equation} \label{eqn:pole-action}
    I_{Pole} = \frac{\beta L}{\pi \lambda} \left( 1 - \sqrt{\frac{c\pi^2  \lambda}{3L^2}-1} \right)=\frac{\beta L}{\pi \lambda} \left( 1 - \sqrt{\frac{c\pi^2  y}{3}-1} \right) = \beta E_{vac}(L)
\end{equation}
for the pole patch.
Note that, when the pole patch can be embedded in global $dS_3$, \eqref{eqn:beta-y-reln} gives us $I_P = \frac{\beta L (1 - \beta/L)}{\pi \lambda}$.

The Cosmic Horizon patch captures the other side of the Dirichlet wall. Here we use a different coordinate notation ($r\to\rho$) to distinguish our two cases.

\[
    ds^2 = \left( \frac{\ell^2 - \rho^2}{\ell^2} \right) d\tau_E^2 + \left(\frac{\ell^2}{\ell^2-\rho^2} \right) d\rho^2 + \rho^2 d\phi^2
\]
with
\begin{equation}
    \rho \in \left[ \rho_c ,\ell \right],\quad \tau_E\sim \tau_E+\beta_c,\quad \phi\sim \phi + 2\pi
\end{equation}
where
\begin{align} \label{eqn:CH-rc}
    \rho_c = \ell\sqrt{\frac{3}{c\pi^2 y}}=\frac{L}{2\pi}, \qquad \beta_c = \frac{\beta}{\sqrt{1-\frac{\rho_c^2}{\ell^2}}}\,.
\end{align}
For the CH patch, the extrinsic curvature at $\rho=\rho_c$ has different signs from \eqref{eqn:pole-ext-curv}:
\begin{align}
    K_{{\tau_E}{\tau_E}} = \frac{\rho_c}{\ell^2}&\sqrt{1-\frac{\rho_c^2}{\ell^2}},\quad  K_{\phi\phi} = -\rho_c \sqrt{1-\frac{\rho_c^2}{\ell^2}},\quad K =  \frac{-1+\frac{2\rho_c^2}{\ell^2}}{\rho_c\sqrt{1-\frac{\rho_c^2}{\ell^2}}} \label{eqn:ch-ext-curv}\\
    &\sqrt{g}\left(K - \frac{1}{\ell}\right) = -1 + \frac{2\rho_c^2}{\ell^2} - \frac{\rho_c}{\ell}\sqrt{1-\frac{\rho_c^2}{\ell^2}}\,.
\end{align}
The first term in \eqref{eqn:eucl-grav} for the CH patch is
\begin{equation}
    -\frac{1}{16\pi G_N} \int_0^{\beta_c}d\tau_E \int_0^{2\pi}d\phi \int_{\rho_c}^\ell d\rho \, \frac{4\rho}{\ell^2} = -\frac{\beta_c}{4G_N}\frac{\ell^2 - \rho_c^2}{\ell^2}
\end{equation}
and the second term is
\begin{equation}
    -\frac{1}{8\pi G_N} \int_0^{\beta_c} d\tau_E \int_0^{2\pi}d\phi \sqrt{g}\left(K-\frac{1}{\ell}\right) = -\frac{\beta_c}{4G_N}\left( -1 + \frac{2\rho_c^2}{\ell^2} - \frac{\rho_c}{\ell}\sqrt{1-\frac{\rho_c^2}{\ell^2}} \right)\,.
\end{equation}
Adding the two terms,
\begin{equation}
    I_{CH} = \frac{\beta_c}{4G_N} \left( -\frac{\rho_c^2}{\ell^2} + \frac{\rho_c}{\ell}\sqrt{1-\frac{\rho_c^2}{\ell^2}} \right) = \frac{\beta \rho_c}{4G_N\ell}\left(1- \sqrt{\left(1-\frac{\rho_c^2}{\ell^2}\right)^{-1}-1}\right)\,.
\end{equation}
Again, using the dictionary \eqref{eq:param-dictionary} and \eqref{eqn:CH-rc} with the relation $(\beta/L)^2 = y/y_0 -1$, we get:
\begin{equation}\label{eqn:CH-action}
    I_{CH} = \frac{\beta L}{\pi \lambda}\left(1-\sqrt{\frac{c\pi^2 y}{3(\beta/L)^2}-1} \right) =\frac{\beta L}{\pi \lambda}\left(1-\sqrt{\frac{c\pi^2 \lambda}{3\beta^2}-1} \right) = L E_{vac}(\beta).
\end{equation}
{where we used the relation:
\[
\left(1-\frac{\rho_c^2}{\ell^2}\right)^{-1} = \frac{y}{y-y_0} = \frac{y}{y_0(\beta/L)^2}.
\]}
Again, in the case in which the bulk can be embedded in global $dS_3$, we find by \eqref{eqn:beta-y-reln} that $I_{CH} = \frac{\beta L (1 - L/\beta)}{\pi \lambda}$.

From \eqref{eqn:pole-action} and \eqref{eqn:CH-action}, we can see that the Pole patch dominates for $\beta > L$ for which $(I_{Pole} < I_{CH})$, and the CH patch dominates otherwise.

\section{Proof of relationship \eqref{eq:Z-with-c6}} \label{app:vac-and-c6}
To prove \eqref{eq:Z-with-c6}, we first recall that
\begin{align}
    S_{Cardy}(\Delta=c/6) &=2\pi\sqrt{\frac{c}{3}\left(\frac{c}{6} -\frac{c}{12}\right)}= \frac{\pi c}{3} = \frac{1}{\pi y_0} \nonumber\\
    -\beta E_{\Delta =c/6}(L) &= -\frac{\beta L}{\pi\lambda}\left(1+\sqrt{-1+\frac{\pi^2 c}{3} y}\right) = -\frac{\beta/L}{\pi y} \left(1+\sqrt{-1+\frac{y}{y_0}}\right) \nonumber \\
    -L E_{vac}(\beta) &= -\frac{\beta L}{\pi\lambda}\left(1-\sqrt{-1+\frac{\pi^2 c y}{3}\frac{L^2}{\beta^2}}\right) = -\frac{\beta/L}{\pi y} \left(1-\sqrt{-1+\frac{y}{y_0}\frac{L^2}{\beta^2}}\right)\,. \nonumber
\end{align}
Using \eqref{eqn:beta-y-reln}, we can rewrite these as:
\begin{equation} \label{eq:rel-vac-c6}
    \begin{aligned}
     S_{Cardy}(\Delta=c/6) & = \frac{1}{\pi y}\,\frac{y}{y_0}=\frac{1}{\pi y} \left( 1+ \frac{\beta^2}{L^2}\right)\\
    -\beta E_{\Delta =c/6}(L) &= -\frac{\beta/L}{\pi y} \left( 1+\frac{\beta}{L}\right)  \\
    -L E_{vac}(\beta) &=  -\frac{\beta/L}{\pi y} \left(1-\frac{L}{\beta}\right) =\frac{1}{\pi y} \left(1- \frac{\beta}{L}\right)
    \end{aligned}
\end{equation}
Now, from \eqref{eq:rel-vac-c6}, we have that
\begin{equation}
    S_{Cardy}(\Delta =c/6) -\beta E_{\Delta=c/6}(L) = \frac{1}{\pi y} \left(1+\frac{\beta^2}{L^2} -\frac{\beta^2}{L^2} - \frac{\beta}{L} \right ) = \frac{1}{\pi y} \left(1- \frac{\beta}{L}\right) = -LE_{vac}(\beta),
\end{equation}
which is \eqref{eqn:beta-y-reln}.

\bibliographystyle{JHEP}
\bibliography{refs.bib}
\end{document}